\documentclass{article}
\usepackage[colorlinks=true, allcolors=blue]{hyperref}
\usepackage[utf8]{inputenc}
\usepackage[symbol]{footmisc}
\usepackage{graphicx}
\usepackage{enumitem}
\usepackage{xcolor}
\usepackage{fullpage}
\usepackage{authblk}
\usepackage{caption}
\usepackage{algorithm2e}[ruled]

\usepackage{amsmath, amsfonts, amsthm, amssymb, dsfont}
\usepackage{thmtools, thm-restate}
\usepackage{braket}
\usepackage{makecell}

\theoremstyle{definition}

\newtheorem*{definition*}{Definition}

\def\correspondingauthor{\footnote{\href{mailto:andrea.gentile@pasqal.com}{andrea.gentile@pasqal.com}}}

\title{Differential equation quantum solvers: \\ 
engineering measurements to reduce cost}

\author[1]{Annie Paine}
\author[1]{Casper Gyurik}
\author[1]{Antonio A. Gentile\correspondingauthor{}}
\affil[1]{{\small PASQAL, 7 rue L\'{e}onard de Vinci, 91300 Massy, France}} 

\date{}

\begin{document}


\maketitle
\setcounter{footnote}{1}

\begin{abstract}

Quantum computers have been proposed as a solution for efficiently solving non-linear differential equations (DEs), a fundamental task across diverse technological and scientific domains.
However, a crucial milestone in this regard is to design protocols that are hardware-aware, making efficient use of limited available quantum resources.
We focus here on promising variational methods derived from scientific machine learning: differentiable quantum circuits (DQC), addressing specifically their cost in number of circuit evaluations. 
The latter ones depend on a factor scaling as $\mathcal{O}(Lm)$ at best, where $L$ is the number of training steps and $m$ is the number of test points at which the trial functions (and their derivatives) must be estimated. 
Reducing the number of quantum circuit evaluations is particularly valuable in hybrid quantum/classical protocols, where the time required to interface and run quantum hardware at each cycle can impact the total wall-time much more than relatively inexpensive classical post-processing overhead.
Here, we propose and test two sample-efficient protocols for solving non-linear DEs, achieving exponential savings in quantum circuit evaluations.
These protocols are based on redesigning the extraction of information from DQC in a ``measure-first" approach, by introducing engineered cost operators similar to the randomized-measurement toolbox (i.e. classical shadows).
Specifically, when adopting a trainable observables model we achieve a scaling factor in circuit evaluations of $\mathcal{O}(dm)$, where the model's expressivity is tuned via a hyperparameter $d \in \text{poly}(n), \, d \ll L$ ($n$ being the number of qubits in the circuit), 
whilst manipulating the cost operator to map the relevant features of the differential equations leads to a scaling factor of $\mathcal{O}(L\log(m))$.
In benchmark simulations on one and two-dimensional DEs, we report up to $\sim100$ fold reductions in circuit evaluations. 
Our protocols thus hold the promise to unlock larger and more challenging non-linear differential equation demonstrations with existing quantum hardware.

\end{abstract}

\section{Introduction}
\label{sec:intro}

Differential equations (DEs) are a cornerstone in the modeling of a vast range of natural phenomena, from physics to engineering and beyond: henceforth, solving DEs efficiently has profound implications for various scientific disciplines. 
In this regard, quantum algorithmic research has led to the development of variational approaches, which aim to solve (non)linear DEs, by leveraging near-term quantum devices to generate adaptive, approximate solutions~\cite{lubasch2020variational, Childs2020, kyriienko:de}.

We focus particularly on the latter paradigm, Differentiable Quantum Circuits (DQC, \cite{kyriienko:de}), inspired by recent classical techniques aiming to embed physics formalism in the training process of neural networks~\cite{raissi2019physics}, which undergo the name of Scientific Machine Learning (SciML).
DQC utilizes quantum circuits to evaluate trial functions and their derivatives with respect to the input parameters, in a discrete sample of points within the equation domain (the ``collocation points'').  
DQC holds a promising amenability of this approach for NISQ implementations, and whilst the problem of extracting meaningful classical information from \textit{trained} circuits has been recently addressed~\cite{williams2024addressing}, yet some practical challenges in experimental \textit{training} realizations remain. 
One of them is the number of quantum circuit evaluations. 
Two key factors of the scaling of this value are the number of training steps, and the number of points in the training set. 
Both can scale prohibitively under certain settings, an issue that becomes even more pronounced for quantum hardware where operations such as initialization and resetting, along with gate-speed, accrue wall-time and can be resource intensive, impacting the final benchmark in application-oriented settings~\cite{wack2021quality}.

In this work, we introduce a novel set of trial functions that aim to mitigate this issue during training. 
These new trial functions are engineered in such a way that the corresponding loss function can be minimized using a significantly smaller number of quantum evaluations. 
One proposed approach leverages a flipped model for trial functions~\cite{jerbi:shadows}, offering exponential savings in quantum resources, especially when applied to problems requiring fine-grained discretization or large-scale training sets (as expected for instances that are challenging when adopting traditional classical approaches, e.g. finite element methods).
Alternatively, we propose to adopt explicitly parameterized, trainable observables, ensuring that the required quantum computation is performed only once \textit{before} training, eliminating the dependence of the number of circuit evaluations on the number of training steps.
An overview of the protocols discussed in this manuscript, including their trial functions and associated training costs, is provided in Table~\ref{table:overview}, and we conclude the paper with a series of experiments to demonstrate the effectiveness of the proposed trial functions on a range of differential equations.

\renewcommand{\arraystretch}{1.5}
\begin{table}[]
\begin{tabular}{c|c|c}
\textbf{Name} & \textbf{Trial function} & \makecell{\textbf{Quantum circuit}\\ \textbf{evaluations for training}}\\\hline
Original protocol (Section~\ref{subsec:protocol_vanilla}) & $f_{\alpha}(x) = \langle 0 | \hat{\mathcal{U}}^\dag(x) \hat{\mathcal{V}}^\dag(\alpha) C \hat{\mathcal{V}}(\alpha) \hat{\mathcal{U}}(x) |0 \rangle$ & $Lm$ \\
Trainable observable (Section~\ref{subsec:protocol_training}) & $f_{\alpha}(x) = \langle 0 | \hat{\mathcal{U}}^\dag(x) C(\alpha) \hat{\mathcal{U}}(x) |0 \rangle$ & $dm$\\  
Flipped model (Section~\ref{subsec:protocol_grid}) & $f_{\alpha}(x) = \bra{0}\hat{\mathcal{U}}^\dagger(\alpha)C(x)\hat{\mathcal{U}}(\alpha)\ket{0}$ & $L\log(m)$  \\
\end{tabular}
\caption{Overview of the protocols discussed in this manuscript, including their trial functions and associated training costs. In the trial functions, $\alpha$ represents the trainable parameters of the model, while $x$ denotes the input. 
Additionally, $\mathcal{U}$ and $\mathcal{V}$ correspond to parameterized quantum circuits, $C$ is a fixed observable, and $C(\alpha) = \sum_{i=1}^d \alpha_i C_i$ and $C(x) = \sum_{i=1}^d f_i(x) C_i$ denote observables derived from a set of measurement operators $\{C_j\}_{j=1}^d$. 
In the number of quantum circuit evaluations, $m$ denotes the number of collocation points, $L$ is the number of training steps, and $d$ is the number of observables included in the parameterized observable.}
\label{table:overview}
\end{table}

\section{Methodology}
\label{sec:protocols}

In this Section, we first outline the original DQC protocol~\cite{kyriienko:de}, clarifying the dependency of quantum circuit evaluations required upon key implicit and explicit factors.  
Next, in Sect.~\ref{subsec:protocol_training} \& \ref{subsec:protocol_grid}, we present in detail our novel protocols, achieving respectively independency against the required training epochs, or a reduced dependency in the number of collocation points. 
Even if the techniques we present in this work are in principle applicable to a wide range of variational quantum algorithms (VQAs), we will focus here on the implications under the aforementioned DQC framework. 

\subsection{Variational protocols for DEs: a succinct summary}
\label{subsec:protocol_vanilla}

We define VQAs as algorithms employing parameterized quantum circuits~\cite{cerezo:variational} to represent a trial function $f_{\alpha}(\mathbf{x})$ in the (multi-dimensional) domain $\mathcal{X}$ of the variable $\mathbf{x}$. The latter is supposed to be tuned, in order to ultimately represent a targeted $f (\mathbf{x})$, interpretable as the solution to a particular problem of interest. 
VQAs typically consider trial functions of the form:
\begin{align}
\label{eq:trail_original}
    f_{\alpha}(x) = \langle 0 | \hat{\mathcal{U}}^\dag(x) \hat{\mathcal{V}}^\dag(\alpha) C \hat{\mathcal{V}}(\alpha) \hat{\mathcal{U}}(x) |0 \rangle,
\end{align}
whose circuit parameters $\alpha$ can be optimised to minimise a given loss function, capturing the distance between $f_{\alpha}(\mathbf{x})$ and $f (\mathbf{x})$. $\hat{\mathcal{U}}(x)$ captures the dependence upon  $\mathbf{x}$ and is hence often called the ``feature map''.
The only quantum measurements required are evaluations of meaningful observables $C$, at the output of these parameterized quantum circuits.  Examples of algorithms which fulfil these requirements can be found in~\cite{cerezo:variational}.
 
We choose to demonstrate our techniques by exemplifying their application to a special case of VQAs solving differential equations (DEs): the DQC protocol. 
The full details for DQC are developed and explained in \cite{kyriienko:de}, however we provide here a brief overview to enhance readability.
Consider a differential equation of the form:  
\begin{align}
    DE_j(\mathbf{x}, \mathbf{f}, \mathbf{df}) = 0,
\end{align}
where $\mathbf{x}$ represents the independent variables, $\mathbf{f}$ the dependent variables, and $\mathbf{df}$ their derivatives with respect to $\mathbf{x}$. 
In order to simplify notation, in the following we will refer to one-dimensional variables $x$, preserving the full generality of the considerations for any multidimensional problem dependent on $\mathbf{x}$ of arbitrary dimensionality. 
To solve this system of equations, trial functions $f_\alpha(x)$ of the form in Eq.~\eqref{eq:trail_original} are utilized.
In this framework, $x$ (the independent variables) are encoded as a continuous parameter in the quantum gates, while $\alpha$ (the variational parameters) represent a set of tunable gate parameters. This allows the model to access the trial function $f_\alpha(x)$ continuously with respect to $x$. 
Moreover, provided that certain requirements on the parameterized gates are met (i.e. the number of distinct eigenvalues of the generators of the parameterized gates is small), derivatives of $f_\alpha(x)$ with respect to both $x$ and $\alpha$ can be efficiently computed on a quantum computer via (generalised) parameter-shift rules~\cite{schuld2019evaluating, kyriienko:gpsr}.  

To solve the DE, the parameters $\alpha$ are optimized to minimize a loss function of the form:
\begin{align}
    \mathcal{L}_{DE}(\alpha) = \sum_{j,i} L\big(DE_j|_{x_i}, 0\big),
\end{align}
where $L(x, y)$ is a distance measure that is zero when $x = y$ and positive otherwise, and $\{x_i\}_{i=1}^m$ is a set of points where the DE condition is evaluated. These \emph{collocation points} are often placed in a regular grid, but not necessarily so. This loss function is minimized when the trial function $f_\alpha(x)$ satisfies the DE at all training points. Any boundary or initial conditions can also be incorporated into the loss function or through post-processing of the quantum trial functions (as discussed in~\cite{kyriienko:de}).  

The optimization proceeds as follows: an initial set of parameters $\alpha$ is chosen, and the quantum model is evaluated at each training point.
If a gradient-based optimizer (e.g., Adam, LBFGS) is used, derivatives of the loss with respect to $\alpha$ are computed using the parameter-shift rule. 
The optimizer updates $\alpha$ iteratively, and this process is repeated for each epoch, until the loss converges or another stopping criterion is met. 
Once training is complete, the converged trial function $f_\alpha(x)$ provides a representation of the DE solution. 
Importantly, because $x$ is continuously encoded in the quantum circuit, the solution can be accessed at any point, not just the training points.

Crucially, for this approach the number of training steps $L$ and the number of collocation points $m$ influences the scaling of the total number of quantum circuit evaluations as:
\begin{align}
    N_{\text{eva}} \propto Lm, 
\end{align}
where $L$ is the number of training steps and $m$ is the number of collocation points. As mentioned in Sect.~\ref{sec:intro}, this can be unfavourable in some situations. Indeed, $L$ can be expected to increase heuristically with the complexity of the problem, as the corresponding optimization problem encounters a more complex landscape, and a number of restarts might be necessary to elude local minima. The grid size $m$ can also be intuitively expected to scale geometrically with the dimensionality of the problem to keep constant the density of points ``anchoring'' the parameterised function to the solution, similarly to classical methods. Whilst the promise inherited by DQC from SciML methods is to alleviate or possibly remove the curse of dimensionality inherent in \textit{scale resolving}, and more specifically grid-based methods~\cite{meng2020ppinn}, it is yet unclear to what extent such alleviation holds, and whether it can be invoked to make them favourable in performance against traditional methods under the most general settings~\cite{grossmann2024can}.

Based upon such reasoning, we will consider in the following that scaling in both $L, m$ factors can be an important factor in determining the (non) implementability of DQC methods in realistic settings.
Therefore, we address such problem in the remaining of this section by introducing two families of quantum trial functions, aimed at (exponentially) improving the circuit-evaluation efficiency of the DQC protocol with respect to either the number of training steps $L$, or the dependence on the $m$ grid points.

\subsection{Trainable observable model}
\label{subsec:protocol_training}

As shown in the previous section, the original protocol requires all function evaluations to be repeated for each epoch, leading to $N_{\text{evo}} \propto L$. 
How the number of training steps scales for a given problem is generally unknown, but large number of required training steps is believed to be likely, due to non-convex landscapes. 
Therefore, we consider an approach where quantum resources are utilized \emph{pre- and post-training} but not \emph{during} training.
Towards this purpose, we consider a trial function of the form
\begin{align}
    f_{\alpha}(x) &= \langle 0 | \hat{\mathcal{U}}^\dag(x) C_\alpha \hat{\mathcal{U}}(x) |0 \rangle, \label{eq:epoch_qmod}\\
    C_\alpha &= \sum_{j=1}^d \alpha_j C_j,
\end{align}
where $\{C_j\}_{j=1}^d$ is a set of measurement operators. 
In such a model we notice that all required function and relevant derivative values can be constructed (no matter the $\alpha$ values) from the following set of measurements:
\begin{align}
    c^{(k)}_{i,j}(x) = \frac{d^k}{dx^k}\langle 0 | \hat{\mathcal{U}}^\dag(x) C_j \hat{\mathcal{U}}(x) |0 \rangle,
    \label{eq:obs_epoch_qmod}
\end{align}
for a given $x$. 
These circuit measurements can be evaluated as in the original protocol (i.e. with parameter shift rule for derivatives), but remarkably, each of these measurements requires a circuit which is independent of the variational parameters. Therefore, the estimate in Eq.~\eqref{eq:epoch_qmod} does not need to be remeasured throughout training.
Consequently, provided the set $\{x_i\}_{i=1}^m$ required for evaluation is known before training, all measurements required for training can be evaluated pre-training. We observe here that fixing the grid beforehand does limit some training strategies - such as variationally altering the grid~\cite{wang20222} - but this restriction can be acceptable under a majority of circumstances.

Adopting the trial function \eqref{eq:epoch_qmod} achieves the goal of removing the dependency upon training steps from the scaling of the required number of quantum measurements. Additionally, measurements made for this epoch sample-efficient method (that we herafter name for simplicity \emph{trainable observable} - TO) can be reused for additional training runs, generalizations of the problem at hand, or multiple functions in the same problem (provided the same $x$-encoding and quantum model are used). 

In a sense, we have effectively moved the variational block from the circuit layers to the measurement operator. When comparing our novel trial functions in Eq.~\eqref{eq:epoch_qmod} with those of the original protocol in Eq.~\eqref{eq:trail_original}, we notice how the measurement operator can be interpreted as $C_{\alpha} \equiv \hat{\mathcal{V}}_\alpha^\dag C \hat{\mathcal{V}}_\alpha$, and consequently think of the \emph{measurement as variational} instead of the state preparation circuit. 
We note that variational cost operators have previously been considered to improve the usual training scheme \cite{kyriienko:de, chen2025learning}, whereas here we investigate adopting them as a full replacement of the variational ansatz.

\subsubsection{Choice of measurement operators}

We have left open the very important choice concerning the set of measurement operators $\{C_j\}_{j=1}^d$. 
This choice will greatly affect the expressivity and trainability of our model. At one extreme we can chose a full $4^N$ size, linearly independent set which spans the full space of Hermitian operators. 
This would maximise the expressivity of our model but limit trainability due to the exponential parameter space to explore. 
At the other extreme, too limited a choice could result in a quantum model with reduced expressivity: though easily trainable, such a model may be efficiently classically simulable (consequently limiting quantum usefulness), or plainly insufficiently expressive to capture the solution to the problem. 
There are many different possibilities regarding how to choose the set of measurement operators. For example, they could be chosen randomly, or guided by suitability for a given hardware. 

Here we suggest as a natural choice to adopt a subset of Pauli strings of length $N$, with $N$ the number of qubits. 
One viable possibility is to consider $k$-local Pauli strings. The total number of such Pauli-strings for contant $k$ is polynomial in number of qubits - a positive seeing as our quantum measurement requirement scales linearly with the size of $\{C_j\}_{j=1}^d$. 
Additionally, by choosing local observables for measurement, it has been shown that the onset of barren plateaus is reduced as compared to global measurements \cite{cerezo2021cost}. 

For sanity-checking the $k$-local restriction, in this paper we also analyse the full exponential $4^N$ set of Pauli strings of length $N$. 
The linear sum of this set $C_\alpha = \sum_j \alpha_j P_j$ spans the full set of Hermitian operators - maximizing the expressivity of our TO model. 
However, this latter option cannot be adopted as a default. As we discuss in detail below, we traded the scaling dependence in quantum measurements on number of epochs with a scaling depending on terms in the cost operator. 
If the latter becomes exponential in $N$, as it is the case for the full set of Pauli strings, then the method would be inefficient. 
Additionally, as each term in the cost operator has a corresponding training parameter, the number of training parameters is now also exponential. 

In conclusion, even when restricting to Pauli strings, which subset of them to choose impacts directly the method performance and requires careful compromise to rule out alternatives. In some situations (and particularly with small circuits), we envisage how the exponential number triggered by the full set of Pauli strings could be yet advantageous against the number of epochs needed to train the variational circuit. We will test this numerically in Sect.~\ref{sec:experiments}.

\subsubsection{Cost of the procedure}

Observing Eqs.~(\ref{eq:epoch_qmod}-\ref{eq:obs_epoch_qmod}), it is immediate to infer how the number of quantum measurements for the proposed \emph{training} regime scales as $\mathcal{O}(dm)$, where $m$ is the size of the training grid and $d$ the number of observables forming the chosen cost operator. 
As discussed earlier, one can realistically expect to restrict the latter set of observables to $d = \mathrm{poly}(n)$ in the circuit size $n$ - e.g. by choosing local Pauli strings. 
In comparison, the original approach requires $ \mathcal{O}(Lm)$ circuit evaluations, so our proposed method effectively removes the $ \propto L$ term, replacing it with $d$.
Therefore, a saving on quantum resources required can be found for problems with $L > d$. Even if this is difficult to estimate in advance, we will show in Sect.~\ref{sec:experiments} how it is easy to find such examples. 
Further savings can be found when considering using the same measurements for multiple functions - e.g. multiple different problems or problems concerning multiple dependent variables.

Finally, a word on \emph{inference}. It is common in (Sci)ML to infer the learned $f({x})$, after training, at a new set of points $\{\tilde{x}_k\}$. Generalising beyond the training set is a much desired feature of machine learnable models. 
Inferring the values $f({\tilde{x}_k}) $  requires more quantum measurements for both the TO and original approaches. However, similarly to what occurs for the training collocation points, in TO $d$ more measurements are required per additional $\tilde{x}_k$ point. 
The number of measurements required for the original method depends on the choice of measurement operator, but generally will be lower due to the variational ansatz, and standard instances employ only $1$ measurement for each estimate of $f({\tilde{x}_k}) $ (leaving aside the estimation of any derivatives).
This drawback must be kept in mind, when adopting TO for problems where we expect to sample the solution in a vast number of points, post training.

\subsection{Flipped shadow model}
\label{subsec:protocol_grid}

We recall from Sect.~\ref{subsec:protocol_vanilla} how the DQC protocol invokes a quantum circuit to evaluate the trial function (and its derivatives with respect to the input and variational parameters) for every collocation point $\{x_i\}_{i=1}^m$. 
This results in the term scaling linearly with $m$ in the number of quantum circuit evaluations. With the \emph{flipped shadow} (FS) model introduced in this section, we target ameliorating this dependency.


In this approach, we consider a slightly different family of trial functions of the form
\begin{align}
\label{eq:trial_flipped}
    f_{\alpha}(x) = \bra{0}\hat{\mathcal{U}}^\dagger(\alpha)C(x)\hat{\mathcal{U}}(\alpha)\ket{0}.
\end{align}
Now $\alpha$ is a set of training parameters that characterise a unitary $\hat{\mathcal{U}}(\alpha)$, whereas $C(x)$ is a measurement operator parameterized by the input parameters $x$.
We can interpret the difference between the trial function in Eq.~\eqref{eq:trail_original} and Eq.~\eqref{eq:trial_flipped} as the order in which the parameters are applied. 
Specifically, in contrast to the original trial function, the FS model applies the variational ansatz $\hat{\mathcal{U}}(\alpha)$ first and subsequently lets the input parameter $x$ encode the measurement. 
That is, we ``flip'' the roles of the variational parameter $\alpha$ and the input parameter $x$.
Flipped models were formally studied in~\cite{jerbi:shadows}, showing how under certain conditions on the observable, original and flipped models are equivalent (i.e., for every trial function, there is an equivalent flipped trial function, and vice versa). 

\subsubsection{Training a flipped model and its cost}
\label{sec:train_flipped}

The idea behind choosing the family of trial functions given in Eq.~\eqref{eq:trial_flipped} is to employ classical shadow methods~\cite{huang:nature} to obtain the necessary data about all collocation points for a single optimization step, using a total number of quantum circuit evaluations that scales as $\log(m) + k$.
Using classical shadows (shadow tomography) to perform a more efficient ``quantum backpropagation'' in terms of circuit evaluations was studied in~\cite{abbas:qbackprop} entertaining generic classes of observables. Here instead we focus on a specific observable, and derive conclusions for adopting FS models in differentiable quantum circuits.

Recall that in every epoch, we need to compute $(d^j/dx^j) f_\alpha(x)$ for every $j = 1, \dots, k$ in the order of the DE, and $x_i \in \mathcal{X}$. 
To achieve this, here we prepare $\tilde{\mathcal{O}}((\log(m) + k)/\epsilon)$\footnote{Here $\widetilde{\mathcal{O}}$ is used to hide scaling with respect to $\log(\delta^{-1})$, where $\delta$ denotes the failure probability.} copies of $\ket{\psi(\alpha)} = \hat{\mathcal{U}}(\alpha) \ket{0}$ and use~\cite{huang:nature} to obtain its classical shadow $\widehat{\psi}(\alpha)$.

We choose to utilize Pauli-shadows, which imposes the condition that the cost function $C(x)$ must be of constant locality to allow for these shadows to efficiently estimate the corresponding expectation values~\cite{abbas:qbackprop}.
To ensure shot-efficient differentiability, we additionally impose that there must exist observables $C^{(j)}(x)$ of constant locality such that 
\begin{align}
\label{eq:deriv_expval}
(d^j/dx^j) f_\alpha(x) = \bra{0} \hat{\mathcal{U}}^\dagger(\alpha) C^{(j)}(x) \hat{\mathcal{U}}(\alpha) \ket{0}
\end{align}
for every $j = 1, \dots, k$ and $x_i \in \mathcal{X}$. 
An obvious choice satisfying the above conditions is the family of observables $C(x) = \sum_l g_l(x) P_l$, where $\{g_l(x)\}_l$ is a family of differentiable functions and $\{P_l\}_l$ is a set of local Pauli-strings.

Once obtained the classical shadow $\widehat{\psi}(\alpha)$, we can use classical postprocessing to estimate $(d^j/dx^j) f_\alpha(x)$ up to precision $\epsilon$ for every $j = 1, \dots, k$ and $x_i \in \mathcal{X}$.
To compute $(d^j/d\alpha^j) f_\alpha(x)$, we basically do the same, but appropriately shifting the variational parameter.
Indeed, in the basic case where employing Pauli-rotations $e^{\alpha_k P_\ell}$ to encode $\alpha$, we can use the parameter shift rule and invoke the classical shadows of $\ket{\psi(\alpha \pm \pi/2)}$, whereas for arbitrary generators of the parameterized gate, we can instead adopt the generalized parameter shift rule~\cite{kyriienko:gpsr}.
We summarise the details of the training procedure in Algorithm~\ref{alg:flipped_model}.

In terms of \emph{cost} of the training procedure, the scaling $\log(m) + k$ in quantum circuit evaluations essentially derives from the preparation of sufficient copies of the fiducial state, in order to obtain its classical shadow.
Thus, an exponential savings of the total number of quantum circuit evaluations against the number of evaluated points in the domain can be made.
This can be particularly relevant for applications 
where fine-grained discretization in the domain, or in general a large number of points to train the solution, is essential.

\section{Experiments}
\label{sec:experiments}

In this Section, we present the algorithmic performance achieved when both the trainable observable and flipped shadows methods are applied to various use-cases.
For more details on the experimental settings, including the choice of ansatze $\mathcal{U}$ and $\mathcal{V}$, the set of observables $\{C_j\}$, the selection of the loss function, and the number of qubits, we refer to Appendix~\ref{appendix:methods}.
We first begin in Section~\ref{subsec:odes} by describing the differential equations used to benchmark our protocols, and afterwards in Section~\ref{subsec:analysis} we analyse the experimental results.

\subsection{Benchmark differential equations}
\label{subsec:odes}

\begin{figure}[h!]
    \centering
    \includegraphics[width=\linewidth]{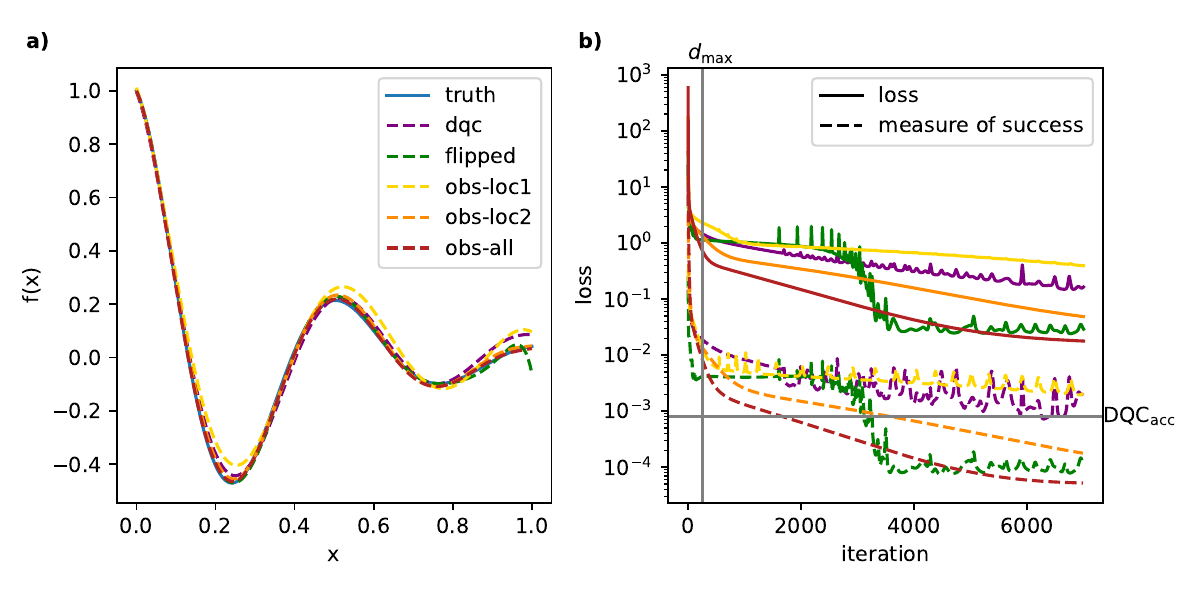}
    \caption{Solving the damped oscillator Eq.~\eqref{eq:damped_osc}. Training as detailed in main text. (a) The solution plotted as attained via DQC, FS and various TO models (dashed lines). ``loc$k$'' TO models include only $k$-local Pauli strings, whereas \emph{all} indicates inclusion of all $4^N$ Pauli strings for the circuit. Exact solution for comparison plotted with a solid line. (b) The loss (solid lines) and measure of success at each iteration of the training for various models. Models identified by same colour as in (a). Solid vertical (horizontal) gray lines indicate as reference the maximum number of circuit executions $d_{\text{max}}$ invoked by any TO model (the baseline MoS performance attained by DQC within the shown iterations). 
    }
    \label{fig:dampedosc}
\end{figure}

We start by considering a simple 1D linear problem corresponding to a damped oscillator:
\begin{align}
    DE(df/dx, f, x) = \frac{df(x)}{dx} + \kappa \exp(-\kappa x) \cos(\lambda x) + \lambda \exp(-\kappa x) \sin(\lambda x) = 0, ~~ f(x_0) = f_0,
    \label{eq:damped_osc}
\end{align}
with $\kappa = 3$, $\lambda = 12$, $x_0 = 0$, and $f_0 = 1$. 
The known, analytic solution of this problem is $\bar{f}(x) = \exp(-\kappa x) \cos(\lambda x)$.
The collocation points here consist of a training grid with $20$ uniformly spaced points in $(0,1)$.
The results of the training are presented in Fig.~\ref{fig:dampedosc}.

\begin{figure}[h!]
    \centering
    \includegraphics[width=\linewidth]{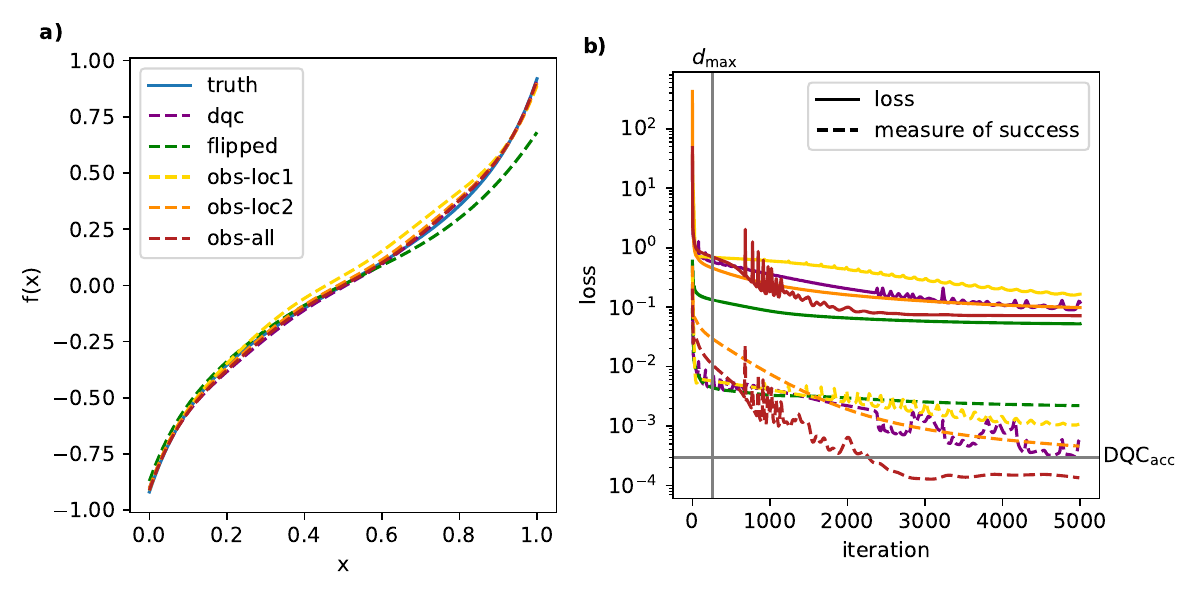}
    \caption{Solving the stationary Burgers Eq.~\eqref{eq:stnry_burgers}. Training as detailed in main text. (a) The solution plotted as attained via DQC, FS and various TO models (dashed lines). Refer to Fig.~\ref{fig:dampedosc} for details. Exact solution for comparison plotted with a solid line. (b) The loss (solid lines) and measure of success at each iteration of the training for various models. Models identified by same colour as in (a). Solid vertical (horizontal) gray lines as in Fig.~\ref{fig:dampedosc}.}
    \label{fig:burgers}
\end{figure}

Secondly, we consider the stationary Burgers equation with two boundary conditions
\begin{align}
    DE(df/dx, f, x) &= f(x)\frac{df(x)}{dx} - \nu \frac{d^2f(x)}{dx^2} = 0, \nonumber\\
    f(x_+) &= f_+ \nonumber \\ 
    f(x_-) &= f_- 
    \label{eq:stnry_burgers}
\end{align}
with $\nu =0.1$.
Burgers equation is a convection-diffusion equation with applications in a range of fields, such as fluid dynamics. 
Compared to the damped oscillator (a linear, first order, 1D problem), stationary Burgers equation is a nonlinear, second order, 1D problem. 
This has solution $\bar{f}(x) = \sqrt{2 \nu a} \tan \left( \sqrt{\frac{a}{2 \nu}} (x+b) \right)$ with $a$ and $b$ two parameters determined by choice of initial condition. 
We choose $f_{\pm 1}$ such that $a = 1$ and $b=0.5$. 
For the training grid, we use again $20$ uniformly spread points within $(0,1)$.
The results of the training are presented in Fig.~\ref{fig:burgers}.

\begin{figure}[h!]
    \centering
    \includegraphics[width=0.9\linewidth]{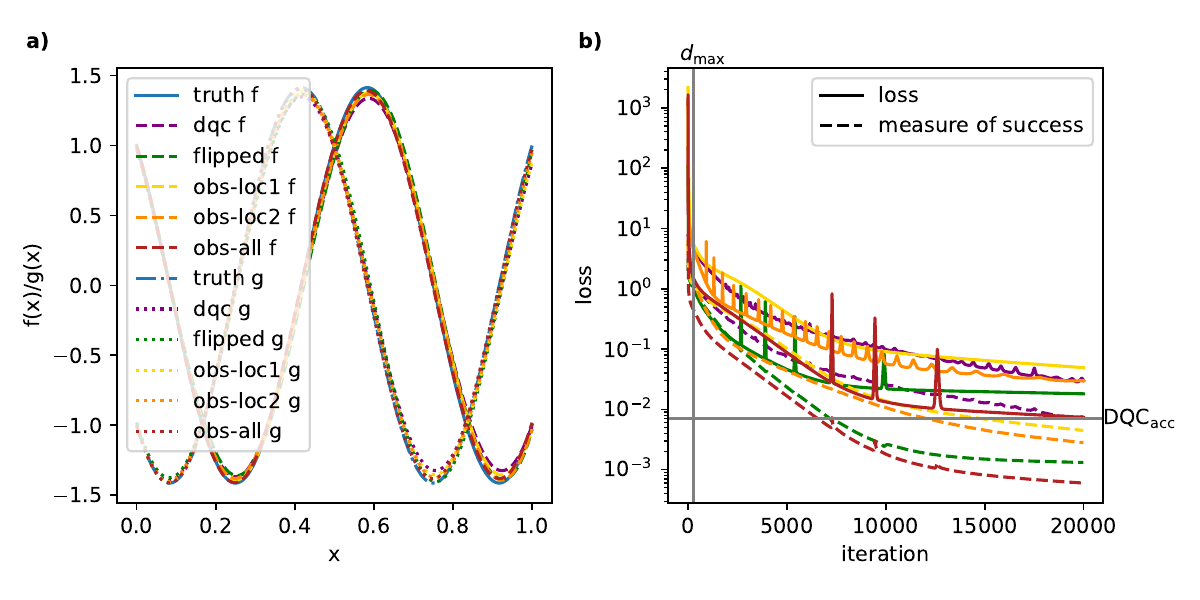}
    \caption{Solving the coupled differential equations Eq.~\eqref{eq:cpd_eqn}. Training as detailed in main text. (a) The solutions $f$ and $g$ plotted, as attained via DQC, FS and various TO models (dashed lines). Refer to Fig.~\ref{fig:dampedosc} for details. Exact solution for comparison plotted with a solid line. (b) The loss (solid lines) and measure of success at each iteration of the training for various models. Models identified by same colour as in (a). Solid vertical (horizontal) gray lines as in Fig.~\ref{fig:dampedosc}.}
    \label{fig:coupled}
\end{figure}

We now consider a pair of coupled equations
\begin{align}
    DE_1(df/dx, dg/dx, f, g x) &= \frac{df(x)}{dx} - 3\pi g(x) = 0, \nonumber \\ 
    DE_2(df/dx, dg/dx, f, g x)&=\frac{dg(x)}{dx} + 3\pi f(x) = 0, \nonumber \\ 
    f(0) &= f_0, ~~ g(0) = g_0
    \label{eq:cpd_eqn}
\end{align}
with $f_0 =1$ and $g_0=-1$. 
This has a solution $\bar{f}(x) = f_0 \cos(3\pi x) + g_0 \sin(3 \pi x )$ and $\bar{g}(x) = -f_0 \sin(3\pi x) + g_0 \cos(3 \pi x )$. 
We stress that the two functions $f(x)$ and $g(x)$ need to be modelled separately. 
For the training grid, we use 20 uniformly-spread points within $(0,1)$.
The results of the training are presented in Fig.~\ref{fig:coupled}.

\begin{figure}[h!]
    \centering
    \includegraphics[width=0.9\linewidth]{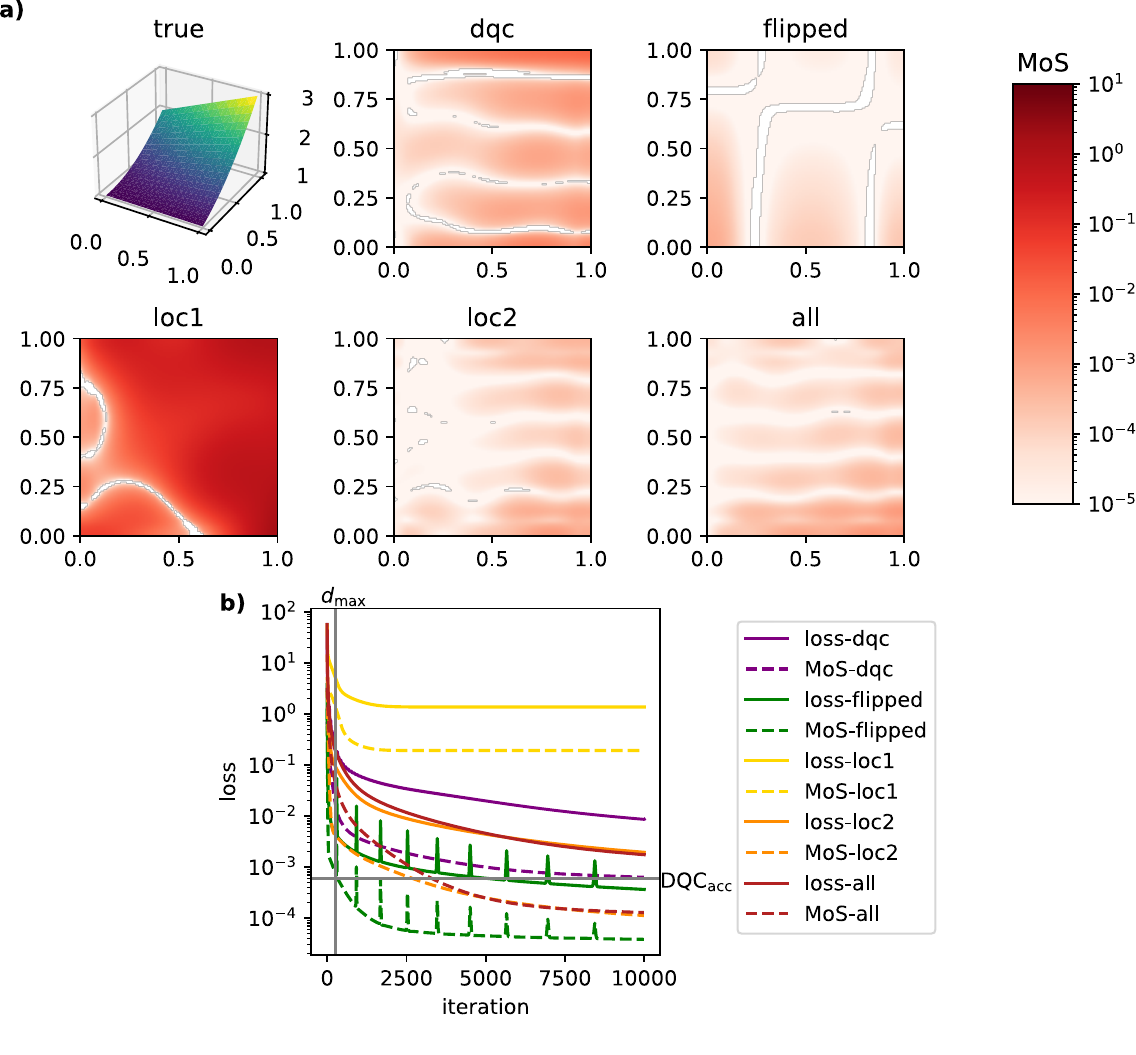}
    \caption{Solving higher dimensional DEs (Eq.~\eqref{eq:hd_ex}). (a) Plots of the known analytic solution (true) and error plots (red-white) of the measure of success (MoS) for various quantum models as introduced in Fig.~\ref{fig:coupled} upon training - as detailed in main text. MoS is here evaluated as the square error between trained solution $f({x}, {y})$ and exact $\bar{f}({x}, {y})$ at each point in the training grid. (b) The loss (solid lines) and MoS (dashed lines) as a function of iteration throughout training for each model. Solid vertical (horizontal) gray lines as in Fig.~\ref{fig:dampedosc}.}
    \label{fig:highdim}
\end{figure}

In our final example, we exemplify higher dimensional problems via the 2D equation:
\begin{align}
    DE(df/dy, x, y)= \frac{df(x,y)}{dy}-2y-x = 0, ~~ f(x, 0) = 1.
    \label{eq:hd_ex}
\end{align}
This problem has known analytic solution $\bar{f}(x,y) = y^2 + xy + 1$. 
For the training grid, we use $20 \times 20$ uniformly-spread points within $(0,1)^2$.
The results of the training are presented in Fig.~\ref{fig:highdim}.

\subsection{Analysis of benchmark results}
\label{subsec:analysis}

For all models, beside visualising the final solution attained, we report the training loss (as defined in Eq.~\ref{eq:loss_terms}) to observe the convergence of all models when the training is halted, as well as the Measure of Success (MoS) as a mean squared error: $\sum_i [f({x}_i) - \bar{f}({x}_i)]^2$ to verify quantitatively how the trained function $f$ fits the exact solution $\bar{f}$ at the collocation points. We note how in sparse grids, unlike those adopted for the experiments here, an additional test set $\{\tilde{x}_i\}$ could be adopted to validate the generalisability of the solution beyond the training set, hence verifying it against overfitting phenomena.  

\paragraph{Trainable observable} 

First, let us discuss the general performance of the TO model against varying numbers of observables entertained. Across the set of examples, we see how adopting \emph{all} Pauli strings performs best, as expected for the most expressive and also most expensive option considered. 
Limiting to $2$-local Pauli strings is often not quite as accurate, but performs on-par with or better than the original trial function under all the exemplified benchmarks, indicating a viable option. 
Finally, $1$-local Pauli strings appear to lack sufficient expressivity for reliable results: if in 1D cases the solution is overall captured but the accuracy is diminished, more complex shapes involved in 2D solutions appear very poorly reproduced (see Fig.~\ref{fig:highdim}).

In terms of savings, whenever the number of epochs considered for the original protocol is greater than the number of observables $d$ entertained, then quantum resource saving is achieved. 
We mark in all figures $d_\mathrm{max}$ as the maximum value of $d$ (that is, considering all Pauli strings, $d = 4^N$). 
As can be seen, in all the examples considered the number of epochs required for the original DQC protocol to converge is well over such threshold, implying that substantially less quantum circuits executions are required. This advantage becomes more and more pronounced as the complexity of the cases increases, as illustrated in Figs.~\ref{fig:coupled}-\ref{fig:highdim} where the corresponding ``saving ratio'' is $\mathcal{O}(10^2)$.

\paragraph{Flipped model} 
The FM consistently achieves a good fit across all benchmarked differential equations, underperforming against the original DQC protocol (for a given number of epochs) only in the case of the Burgers equation, and displaying some outlier points in the simplistic damped oscillator case.  
In terms of efficiency, it reduces the number of quantum circuit evaluations by a factor of $\log(m)/m$, i.e. exponentially growing with the number of collocation points $m$. 
In our experiments, where $m=20 \rightarrow 400$, this resulted in a tenfold to hundred-fold reduction.

\section{Discussion}

We have developed two models specifically targeting a reduction in the quantum resources required to run differentiable quantum models, in particular, focusing on the number of circuit executions. 
Since the proposed TO model removes scaling with respect to number of iterations, whilst the FS model logarithmically reduces evaluations against the size of the training points set, the two methods can act complementarily, and in different scenarios one or the other might become most favourable - as observed in the examples. 

TO models require measuring many Pauli strings for the same state, a topic of interest for various quantum algorithms. 
So far, we limited such cost by restricting to set of $k$-local strings, whilst retaining sufficient expressivity to model all the cases listed in Sect.~\ref{sec:experiments}.
Therefore, works increasing the efficiency of calculating such sets of measurements can be directly embedded to further improve on the method. 
Some of these methods are reviewed in \cite{huang:nature}, with one approach detailed in \cite{chen:memory}. However, we observe how such existing methods come at the cost of (polynomial) increase in qubits and/or probabilistic success chance. Therefore, utilizing such techniques could allow for a wider range of models to be considered efficiently, provided e.g. that larger circuits are an acceptable space-time tradeoff for the experimental platform of choice.

We also note that there is a one-one correspondence between number of training parameters and number of observables for the proposed model. 
Therefore, if large numbers of observables are used, there is a corresponding large number of trainable parameters. 
Large numbers of trainable parameters can result in training difficulties and training overheads. 
We suggest two approaches to reduce this: (1) alternate training regimes and (2) alternate model structures. 
For the first, at each epoch a random polynomial subset of the variational parameters is considered ``trainable" whereas the rest are considered ``static''. With such an approach all parameters are trained but not the full exponential number needs to be considered each epoch. 
With (2) instead, one could adopt any combination of $f_\alpha(x)$, generalising beyond the family of linear combinations of observables considered in this work, without introducing any additional quantum circuit evaluations throughout training. This would alter both expressivity and trainability (in particular the currently enforced 1-1 relation between number of observable and training parameters). For example, a classical neural network taking the measurements as inputs could be considered. 

FS models introduce trial functions designed to invoke classical shadows to reduce the number of circuit evaluations.
In the experiments of Sect.~\ref{sec:experiments}, also FS models displayed a good performance at reproducing the solutions for the various problems modelled.

Whilst already achieving substantial $\mathcal{O} (10 \rightarrow 10^2)$ savings in $N_{\text{eva}}$ in the synthetic experiments reported here, which could be significant in hardware experimental settings, we stress how savings enabled by FS models might become critical in more realistic scenarios. 
Indeed, even if most SciML methods classify as scaling-free, it is yet expected for the number of collocation points $m$ to increase when solving higher-dimensional differential equations. State-of-art examples employing $m \sim 4,000$, like modelling of 3D incompressible flows~\cite{cai2021physics}, would benefit from $\mathcal{O}(10^3)$ reductions in the total number of circuit executions, which can subvert unfeasibility conclusions in some hardware platforms.

Finally, one could adopt neural surrogates to approximate some training data from the quantum computer~\cite{wanner:dl}. 
A neural network architecture can predict mappings of the form $x \mapsto \mathrm{Tr}\left[P \rho(x)\right]$, where $P$ is a local Pauli string and $\rho(x)$ is the ground state of a gapped, parameterized family of Hamiltonians $H(x)$\footnote{Through Kitaev's circuit-to-Hamiltonian construction, output states of constant-depth (parameterized) quantum circuits correspond to the ground states of gapped Hamiltonians.}. 
In our context, this method could be employed by simultaneously training the neural network during the optimization phase. This would allow the network to learn how to approximate $\mathrm{Tr}\left[P \rho(x)\right]$ based on the training data. Once the neural network is trained, it could be used during the deployment phase to estimate $\mathrm{Tr}\left[P \rho(x)\right]$, effectively removing the need for quantum circuit evaluations at that stage.

\subsection{Beyond classical capabilities}

The question of ``why quantum?'' arises often when considering variational algorithms, without formal proof of advantage. 
We stress how in this proposal we retain the same motivation for quantum usage as the original protocol - namely expressivity of the quantum model over its classical counterpart. 

Choosing a TO model does not remove this potential: in other words the proposed model is not generally efficiently classically simulable. For example, choosing $k$-local Pauli-strings as the set of measurement operators leads to trial functions which are generally intractable for any efficient classical model~\cite{molteni2024exponentialquantumadvantageslearning}.
FS models, instead, use parametrized quantum circuits to define a linear combination of basis functions $g_l$ weighting a set of Pauli strings (see Sect.~\ref{sec:train_flipped}). This introduces an inductive bias that is in general difficult to replicate classically. As demonstrated in~\cite{jerbi:shadows}, such a flipped model can efficiently represent—and learn from—functions that cannot be represented classically within a reasonable computational cost.

\section{Final remarks}

This work continues to emphasise just how important the choice of quantum model can be, when utilising quantum variational algorithms. Previous works have emphasised how the choice affects drastically the expressivity \cite{schuld2021effect, du2022efficient} and the trainability \cite{cerezo2021cost, gil2024relation}. How the choice of quantum model can affect quantum resource requirement has also been considered before, such as the choice of hardware efficient ans{\"a}tze \cite{kandala2017hardware}. 

Here, we take a step further in analysing the impact of such choices in quantum resource scaling, and hence effective, practical implementability.
Obvious consequences of our work are to lower the bar in the ease of implementability of increasingly complex showcases of differentiable quantum circuits on real (as well as accurately emulated) near-term quantum devices. 
We also observed how using a trainable observable (TO) model, a set of data can be gathered for a chosen quantum model and training points, and then reused for multiple runs and different problems leveraging upon pure classical post-processing. Similarly, attaining a classical shadow in a flipped shadow (FS) model allows to perform additional operations, generalisation checks, and optimization schedules, all without accruing additional quantum computing wall-time.  

Further work and possible improvements have been mentioned, and we summarise them here. 
The TO model could be generalised beyond linear sums of observables, whereas the FS method could benefit from more advanced tomographic techniques, such as online shadow tomography~\cite{chen2024adaptive}. 
Additionally, more advanced techniques could be considered for Pauli string measurements, applying both to the  TO as well as the FS models. Finally, future work could explore incorporating the methods introduced in \cite{wanner:dl} to further reduce the reliance on a quantum computer to only a handful of crucial steps. 
Other way around, the set of tools deployed here for the specific case of differentiable quantum circuits could be extended to broader variational quantum machine learning settings. 

In conclusion, the strategies analysed here bear the potential to significantly enhance the quantum-resource efficiency of variational quantum algorithms, especially in the considered deployment scenario of solving differential equations, enhancing the implementability of more complex usecases in near-term quantum machines.

\subsection*{Acknowledgments}
The authors acknowledge Pasqal co-workers for useful discussions and manuscript revision, and in particular: Savvas Varsamopoulos, Evan Philip, Raja Selvarajan, Louis-Paul Henry, Mourad Beji as well as Oleksandr Kyriienko for insighftful comments.  
C.G. acknowledges the project ``Divide \& Quantum'' (P.N. 1389.20.241) of the research programme NWA-ORC (partly) financed by the Dutch Research Council (NWO). A.A.G. acknowledges funding via  ``Qu\&Co flow'' project ID (EC): 190146292. 

\bibliographystyle{unsrt}
\bibliography{main}

\appendix

\section{Methods}
\label{appendix:methods}

For our experiments, we used Python to simulate the proposed algorithms, and PyTorch to perform the training. 
Furthermore, we performed noiseless full state simulation using our open sourced tool Qadence~\cite{seitz2024qadencedifferentiableinterfacedigitalanalog}.
Throughout all experiments, and for all models described below, we utilize $N=4$ qubits.

For the flipped model, we select trial functions as given in~\eqref{eq:trial_flipped}, with additional scaling and offset parameters: 
\[
f_\alpha(x) = \alpha_{\mathrm{out}} \langle 0 | \hat{\mathcal{U}}^\dag(\alpha) C(\alpha_{\mathrm{in}}x + \alpha_{\mathrm{shift}}) \hat{\mathcal{U}}(
\alpha) | 0 \rangle + \alpha_{\mathrm{offset}}
\]  
Throughout all experiments we use the Hardware-Efficient Ansatz (HEA) for $\mathcal{U}(\alpha)$. 
Moreover, the observables are $C(x) = \sum_j f_j(x)P_j$, where $\{P_j\}_j$ is the set of $1$-local Pauli strings. 
Depending on the experiment, $f_j(x)$ is either the $j$th monomial (with respect to lexographical ordering) or the $j$-th Chebyshev polynomial.

For the trainable observable model, we select trial functions as given in~\eqref{eq:epoch_qmod}, with an additional scaling parameter:  
\[
f_\alpha(x) = \alpha_s \langle 0 | \hat{\mathcal{U}}^\dag(x) C_\alpha \hat{\mathcal{U}}(x) | 0 \rangle, \quad \hat{\mathcal{U}}(x) = \hat{U}_{\mathrm{b}} \bigotimes_j R_X^j(j x).
\]  
The scaling parameter $\alpha_s$ enables the optimizer to efficiently adjust the overall scale of the trial function. 
The encoding circuit $\hat{\mathcal{U}}(x)$ consists of a tower Fourier feature map (as detailed in~\cite{kyriienko:de}) and a random, static unitary $\hat{U}_{\mathrm{b}}$ that changes the basis and introduces entanglement.
Finally, we compare and analyze three sets of observables $\{C_j\}_{j=1}^d$: (1) the full set of $4^N$ Pauli strings, (2) the set of $3N + 1$ 1-local Pauli strings, (3) the set of $9N(N-1)/2 + 3N + 1$ 2-local Pauli strings.  

Additionally, we contrast these with the original protocol outlined in Sect.~\ref{subsec:protocol_vanilla}, where trial functions are defined as:  
\[
f_\theta(x) = \theta_{\mathrm{sc}} \langle 0 | \hat{\mathcal{U}}^\dag(x) \hat{\mathcal{V}}^\dag(\theta) C \hat{\mathcal{V}}(\theta) \hat{\mathcal{U}}(x) | 0 \rangle + \theta_{\mathrm{sh}},
\]  
with  
\[
\hat{\mathcal{U}}(x) = \bigotimes_j R_X^j(jx), \quad C = \sum_{j=0}^{N-1} \hat{Z}_j, \quad \hat{\mathcal{V}}(\theta) = \mathrm{HEA}.
\]

During training, we employ the following loss function:  
\begin{equation}
    \label{eq:loss_terms}
    \mathcal{L} = \mathcal{L}_{\mathrm{DE}} + \mathcal{L}_{\mathrm{BC}},
\end{equation}
where:  
\[
\mathcal{L}_{\mathrm{DE}}(\alpha) = \frac{1}{m} \sum_j (\mathrm{DE}|_{x_j})^2, \quad \mathcal{L}_{\mathrm{BC}}(\alpha) = (f_\alpha(x_0) - f_0)^2.
\]  
This loss function minimizes the mean squared error of the initial value difference ($\mathcal{L}_{\mathrm{BC}}$) and the residuals of the differential equation ($\mathcal{L}_{\mathrm{DE}}$) when evaluated at the training points $\{x_j\}_{j=1}^m$. Different training grids are considered for the different experiments.

\RestyleAlgo{ruled}
\SetKwComment{Comment}{/* }{ */}
\begin{algorithm}
\caption{Training a flipped model}\label{alg:flipped_model}
$\alpha \gets \alpha_{\mathrm{init}}$\;
\While{training}{
\For{$\alpha, \alpha \pm \pi/2$ \Comment*[r]{$\alpha \pm \pi/2$ only required for grad-based optimization}}{
Prepare $\mathcal{O}((\log(m) + k)/\epsilon)$ copies of $\ket{\psi(\alpha)} = \hat{\mathcal{U}}(\alpha) \ket{0}$ \;
Use~\cite{huang:nature} to obtain the classical shadow $\widehat{\psi}(\alpha)$ \;
}
\For{$j=0,\dots, k$}{
  \For{$x\in \{x_i\}$}{
    \eIf{j=0}{
    Use $\widehat{\psi}(\alpha)$ to evaluate $f_{\alpha}(x)$ \;
    }{
    Use $\widehat{\psi}(\alpha)$ to evaluate $(d^jf/dx^j) f_{\alpha}(x)$ \;
    }
  }
}
perform optimization step: $\alpha \gets \text{update}(\alpha)$
}
\end{algorithm}

\end{document}